\documentclass[11pt,titlepage]{article}
\usepackage{epsfig}
\usepackage{kotex}
\usepackage{graphicx}
\usepackage{latexsym}
\usepackage{amssymb}
\usepackage{amsmath,amscd}
\usepackage{multirow} 
\usepackage{array}
\usepackage{caption}
\usepackage{rotating}
\usepackage{subfig}
\usepackage{color}
\usepackage{natbib}
\usepackage{rotating}
\usepackage{graphicx,lscape}
\usepackage{amsmath, amsthm, amssymb}
\usepackage{graphicx}
\usepackage{epsfig}
\usepackage{graphics,color, graphicx}
\usepackage{latexsym}
\usepackage{amssymb,amsmath,amscd}
\usepackage{fancyvrb,enumerate}
\usepackage{amsmath, amsthm, amssymb}
\usepackage{mathtools}
\usepackage{verbatim} 
\usepackage{afterpage}
\usepackage{dsfont}

\long\def\comment#1{}

\newcommand{\ba}{\mbox{\boldmath $a$}}
\newcommand{\bg}{\mbox{\boldmath $g$}}

\newcommand{\bff}{\mbox{\boldmath $f$}}

\newcommand{\bZ}{\mbox{\boldmath $Z$}}
\newcommand{\bV}{\mbox{\boldmath $V$}}

\newcommand{\bX}{\mbox{\boldmath $X$}}
\newcommand{\bdm}{\begin{displaymath}}
\newcommand{\edm}{\end{displaymath}}

\newcommand{\bpsi}{\mbox{\boldmath $\psi$}}

\newcommand{\bmu}{\mbox{\boldmath $\mu$}}

\title{\bf  Multi-feature Clustering of Step Data using Multivariate Functional Principal Component Analysis}

\author{
{\sc Wookyeong Song and Hee-Seok Oh} \\
Seoul National University\\
Seoul 08826, Korea \\
\\
{\sc Yaeji Lim} \\
Chung-Ang University\\
Seoul 06974, Korea\\
\\
{\sc Ying Kuen Cheung}\\
Columbia University\\
New York 10032, USA\\
\\
}
\date{Draft: version of \today}

\begin{document}

\maketitle


\begin{center}
{\bf Abstract}
\end{center}

\noindent 
This paper presents a new statistical method for clustering step data, a popular form of health record data easily obtained from wearable devices. Since step data are high-dimensional and zero-inflated, classical methods such as $K$-means and partitioning around medoid (PAM) cannot be applied directly. The proposed method is a novel combination of newly constructed variables that reflect the inherent features of step data, such as quantity, strength, and pattern, and a multivariate functional principal component analysis that can integrate all the features of the step data for clustering. The proposed method is implemented by applying a conventional clustering method such as $K$-means and PAM to the multivariate functional principal component scores obtained from these variables. Simulation studies and real data analysis demonstrate significant improvement in clustering quality.
\vskip 5mm


\pagenumbering{arabic}

\newpage

\section{Introduction}\label{ch1}
Along with a growing interest in digital and smart healthcare, studies of physical activity measured using wearable devices are also on the rise. Analysis of personal health record data can provide a concise and meaningful insight into an individual's state of activity, enabling them to provide customized health care services based on personalized data. \cite{Masurier2005} used pedometers to determine the physical activity levels of American youth. \cite{Bassett2010} analyzed the number of daily steps in various demographic subgroups to identify predictors of pedometer-measured physical activity performed by American adults. In recent years, statistical learning methods have used for activity recognition studies. \cite{Shoaib2015} studied the clustering of living activities by analyzing data from smartphones and smartwatches based on a support vector machine and decision trees. \cite{Balli2017} compared the naive Bayes, $k$-nearest-neighbors, logistic regression, Bayesian network, and multilayer perceptron methods in terms of human activity recognition using smartwatch sensor data. 

This study analyzes step count data recorded from a wearable device, {\it Fitbit}, that tracks the wearer's activity. Data used for the analysis are recorded for 21,394 days for 79 users and are collected at one-minute intervals, yielding 1440 epochs per day per individual. In this paper, we want to cluster ``days" based on physical activity information. 

We propose a new clustering method that reflects the vital intraday characteristics of physical activities such as amount, intensity, and pattern. The proposed method consists of two key elements: the composition of new functional variables and a multivariate functional principal component analysis (MFPCA). The construction of the new variables is designed to represent the step data's inherent features, such as quantity, strength, and pattern. The MFPCA applied to the new variables provides low-dimensional MFPC scores so that some conventional clustering methods can be used to the step data analysis. Specifically, we first generate new variables and apply the MFPCA to the new variables. Classical clustering methods such as $K$-means or partitioning around medoids (PAM) \citep{Kaufman1987} are then applied to low-dimensional MFPC scores. 

In the literature, there are numerous clustering methods for multivariate functional data. \cite{Jacques2014} presented a parametric mixture model for multivariate functional data, which uses the multivariate probability density of the principal component vector as a proxy for the density of the original data. \cite{ Chiou2014} investigated a normalized MFPCA and its application to functional clustering. \cite{Bouveyron2016} proposed a discriminative functional mixture (DFM) model that models data into one distinct functional subspace. The FunFEM algorithm was further proposed for inference using the DFM model. \cite{Schmutz2020} suggested clustering multivariate functional data by projecting data into low-dimensional subspaces using an MFPCA and a functional latent mixture model. We emphasize that the proposed method fully reflects the crucial features of step data: discrete (count), high-dimensional, and zero-inflated, which is the crucial contribution that distinguishes the proposed method from the existing methods. 

In our previous work, \cite{Lim2019} introduced input variables for clustering accelerometer data based on a rank-based transformation and thick-pen transformation.
The main difference is that the proposed method in this article considers the amount, intensity, and pattern of the step count data simultaneously for clustering, while \cite{Lim2019} considered the amount and pattern of activity separately. This is a critical extension as daily step data is a complex process defined by more one single feature such as amount and pattern.  
Also importantly, the proposed method is applicable to discrete and zero-inflated data, which are natural attributes of step data.  This is also  a significant improvement of the existing clustering approaches.

The rest of this paper is organized as follows. Section 2 presents the scheme of constructing new variables. In Section 3, the proposed clustering method based on MFPCA and the constructed variables is proposed. Section 4 discusses real data analysis with having the step count data, and Section 5 further performs simulation studies with various test functions beyond step count data to assess the effectiveness of a general clustering method. Concluding remarks are provided in Section 6.  
  
\section{Construction of Multiple Functional Input Variables}\label{ch2}
  
This section introduces three variables generated from the original step data $X(t)$, $t=1, \ldots, T$, i.e., the amount, intensity, and pattern of physical activity. 
    
\subsection{Cumulative Sum Function}
For a given real-valued process $\{ X(t): t \in (0, T)\}$, the cumulative activity up to $t$ is defined as $S(t) := \int_0^t X(u) du$, $t \in (0, T)$. Since the cumulative sum represents the amount of the activity in the step data, $S(t)$ is considered as a functional variable, termed {\em cumulative sum function}.

\begin{figure}
\centering
\includegraphics[width=\linewidth]{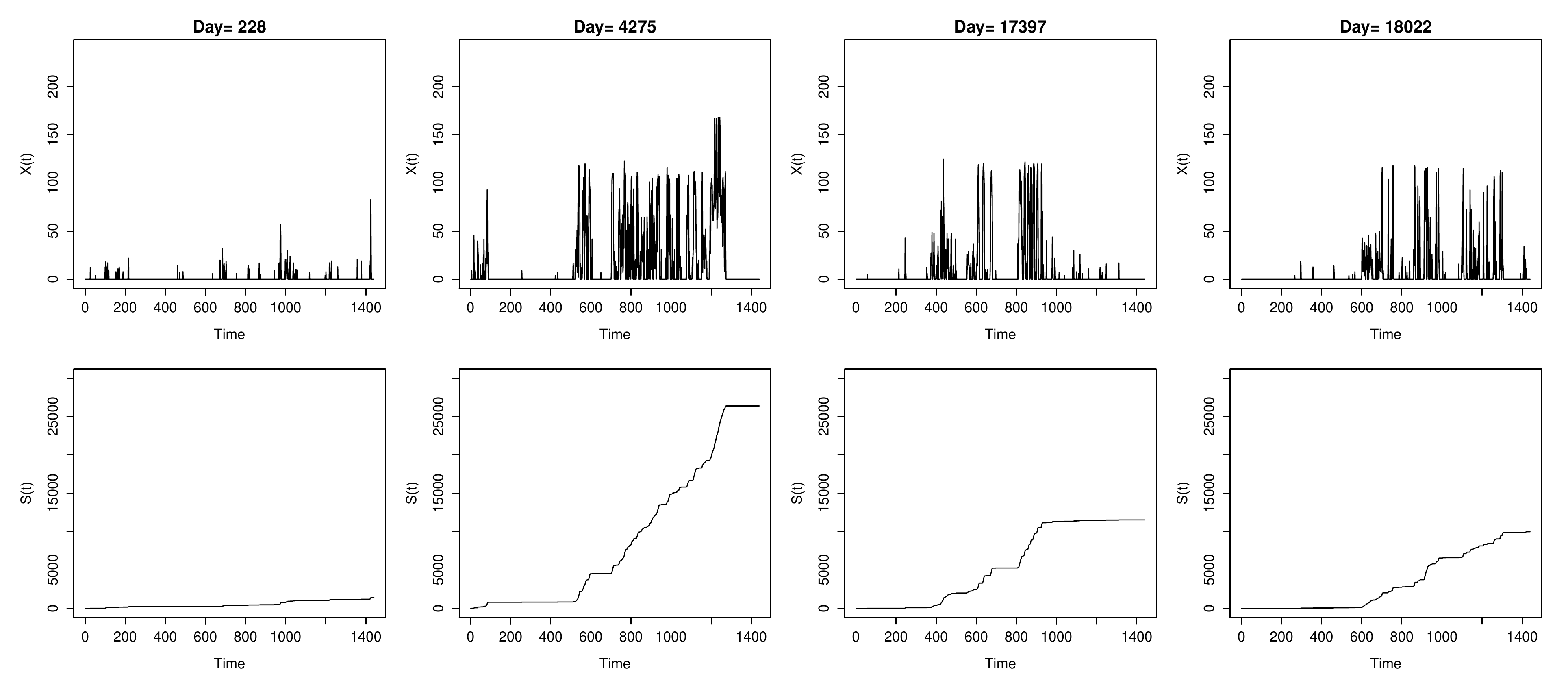}
\caption{Four step count datasets $X(t)$ according to four different days (top row), and the corresponding cumulative sum functions $S(t)$ (bottom row).}
\label{order}
\end{figure}

Figure \ref{order} shows the original step datasets $X(t)$, and the corresponding cumulative sum functions $S(t)$ over the four randomly selected days. By comparing the cumulative sum functions, it can identify the quantity of daily activity. On day 4275, a large amount of activity is observed compared to 228, 17397, and 18022 days. However, focusing on the amount of information in the data may not reflect some vital features like time-related information, such as activity intensity or pattern. For example, the cumulative totals of days 17397 and 18022 seem similar, but these original step data have different patterns.   
  
\subsection{Ordered Quantile Slope Function}
  
\begin{figure}[!h]
\centering
\includegraphics[width=0.75\linewidth]{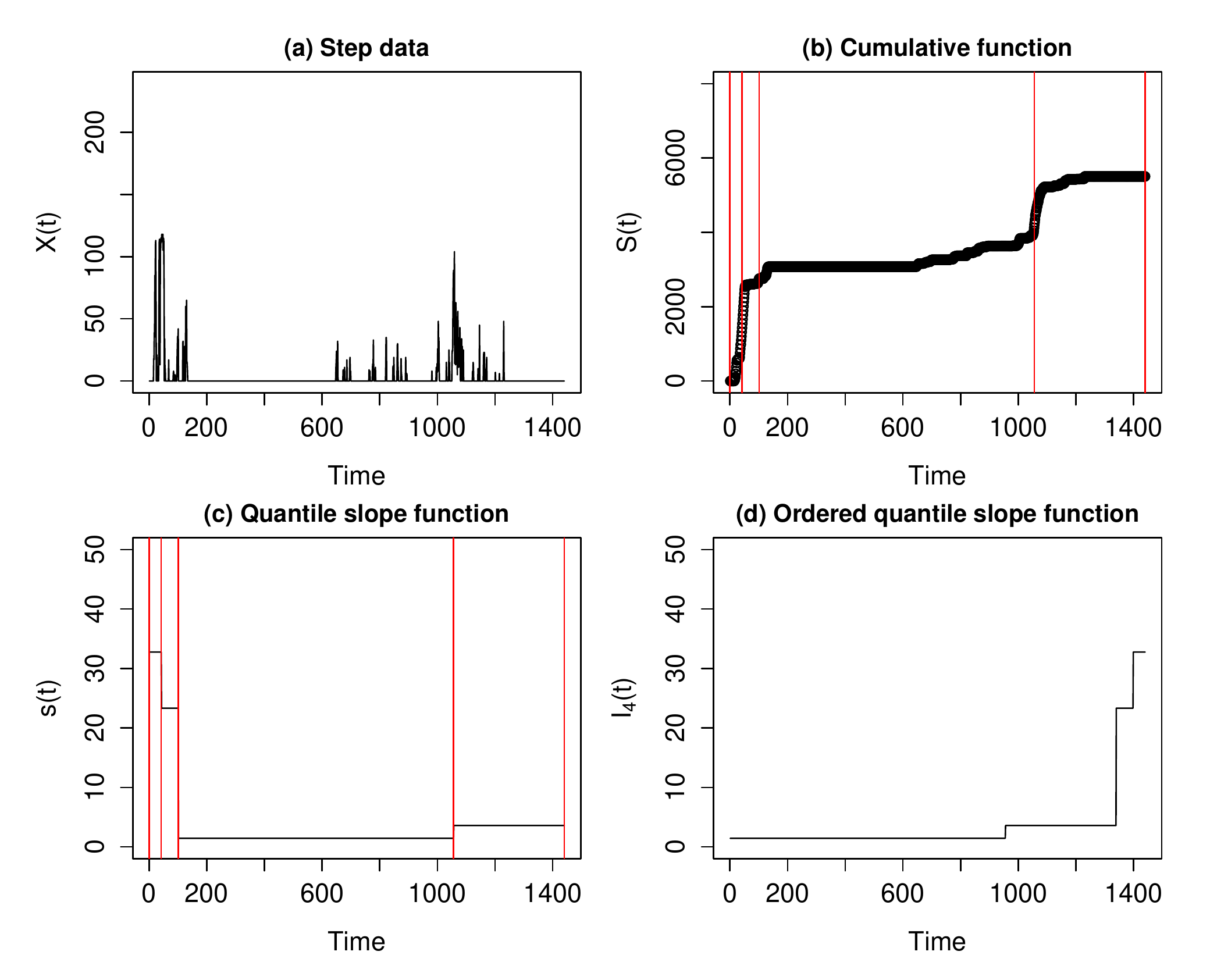}
\caption{(a) Step data $X(t)$ in day 20682, (b) the corresponding cumulative function $S(t)$, (c) quantile slope function $s(t)$, and (d) ordered quantile slope function $I_{Q}(t)$. Note that the red vertical lines indicate quantiles $\mathcal{T} (p_{q})$.}
\label{order_ex}
\end{figure}

The intensity of the step data can be useful for understanding and classifying an individual's state of activity beyond the simple total steps of the data. To this end, a quantile-based function is considered to reflect the intensity of step data \citep{Cheung2018}. We first define the $100p$th quantile of the activity time as 
\begin{eqnarray} 
\mathcal{T} (p) := \inf \{ t | S(t) \geq p S(T) \},
\label{fp}
\end{eqnarray} 
where $S(t) = \int_0^t X(u) du$ as defined in Section 2.1. $\mathcal{T} (p)$ can be interpreted as the time when the $100p$ percent of the total activity has been achieved. For example, $\mathcal{T} (0.5)$ indicates the time to reach the middle activity of the day.  A quantile slope function between $\mathcal{T} (p_q)$ and $\mathcal{T} (p_{q+1})$ is then defined as 
\[
s (t) := \frac{ S(T)/Q } {  \mathcal{T} (p_{q+1})  - \mathcal{T} (p_q)  },~~  \text{ for }   ~~  \mathcal{T} (p_{q})  \leq t < \mathcal{T} (p_{q+1}),
\]
where $p_q = \frac{q}{Q}$, $q=0, \ldots, Q$ and $Q$ is the number of quantiles. The quantile slope function $s(t)$ provides the intensity of the activity, which shows how long it takes to achieve $\frac{1}{Q}$ of the total number of steps per day. Figure \ref{order_ex}(a) and (b) show the step data, $X(t)$ for a particular day and the corresponding cumulative function, $S(t)$. In the figure, the red vertical lines indicate quantiles, $ \mathcal{T} (p_{q})$, where $p_q = \frac{q}{4}$ ($q=0, \ldots, 4$). The quantiles $\mathcal{T} (p_{q})$ seem to detect the high intensity time points well. In addition, the corresponding quantile slope function $s(t)$ is plotted in Figure \ref{order_ex}(c), which shows the slope information in $S(t)$ clearly.   

\begin{figure}[!h]
\centering
\includegraphics[width=\linewidth]{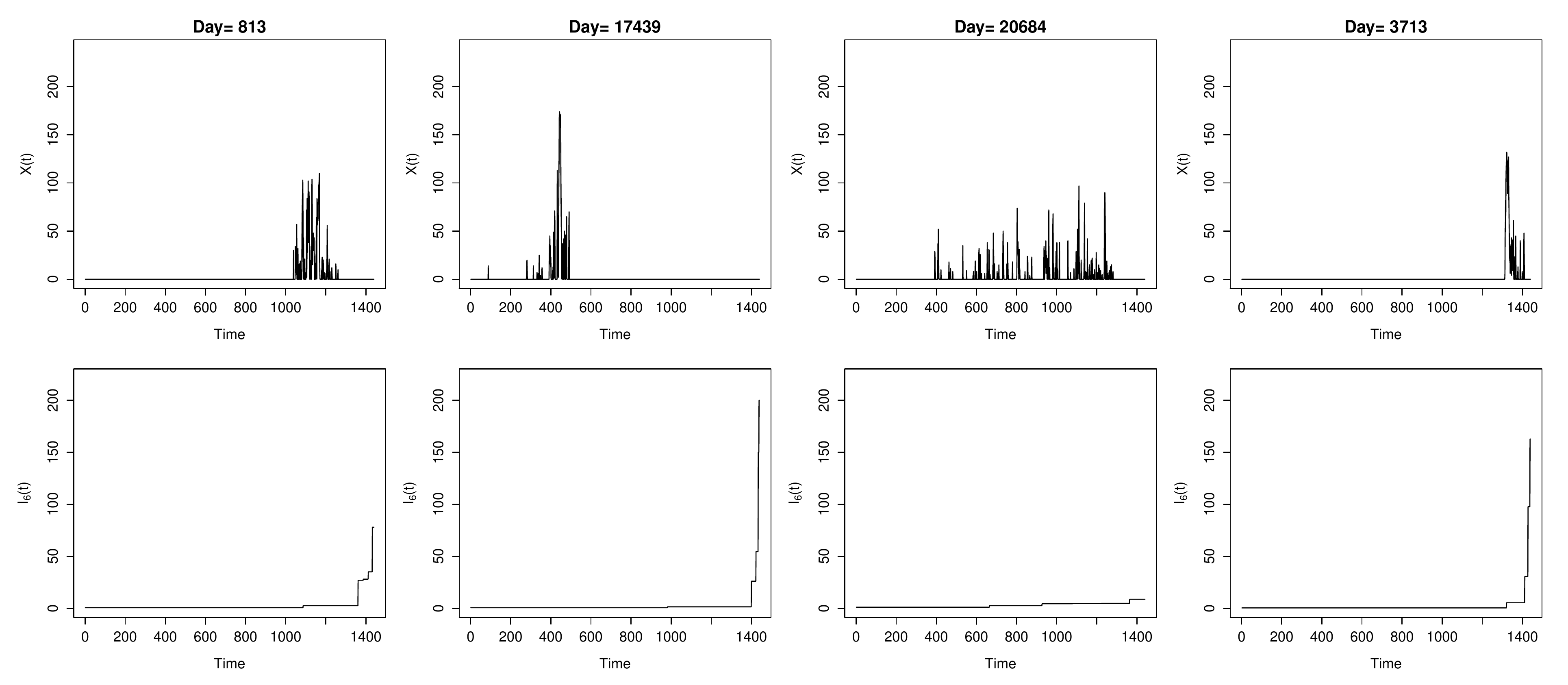}
\caption{Four step datasets $X(t)$ (top row), and the corresponding ordered quantile slope functions $I_{Q}(t)$ with $Q=6$ (bottom row).}
\label{order_slope}
\end{figure}

To examine the intensity of the activity further, we eliminate the time information of $s(t)$ by ordering it, 
$$ 
I_{Q}(t) := s_{(t)}~,~~~ t=1,2, \ldots , T,
$$
where $s_{(t)}$ is $t$th smallest value of $s(t)$. $I_{Q}(t)$ is termed {\em ordered quantile slope function} of $X(t)$ with $Q$ quantiles. Figure \ref{order_ex}(d) shows $I_{Q}(t)$ of step data $X(t)$, and Figure \ref{order_slope} shows $I_{Q}(t)$'s from four randomly selected days. As one can see, the number of step counts on day 17439 is tiny, except near $t=500$, where a spike in the number of steps occurred. Likewise, on day 3713, the surge in activity is about $t=1340$. The ordered quantile slope function $I_{Q}(t)$ shows such intensity at a high peak close to $t=1440$. Meanwhile, the activity on day 813 is concentrated between $t=1000$ and $t=1250$ but is not as intense as that on days 17439 and 3713. Therefore, $I_{Q}(t)$ on day 813 is much lower than that on days 17439 and 3713.

\begin{figure}[!h]
\centering
\includegraphics[width=\linewidth]{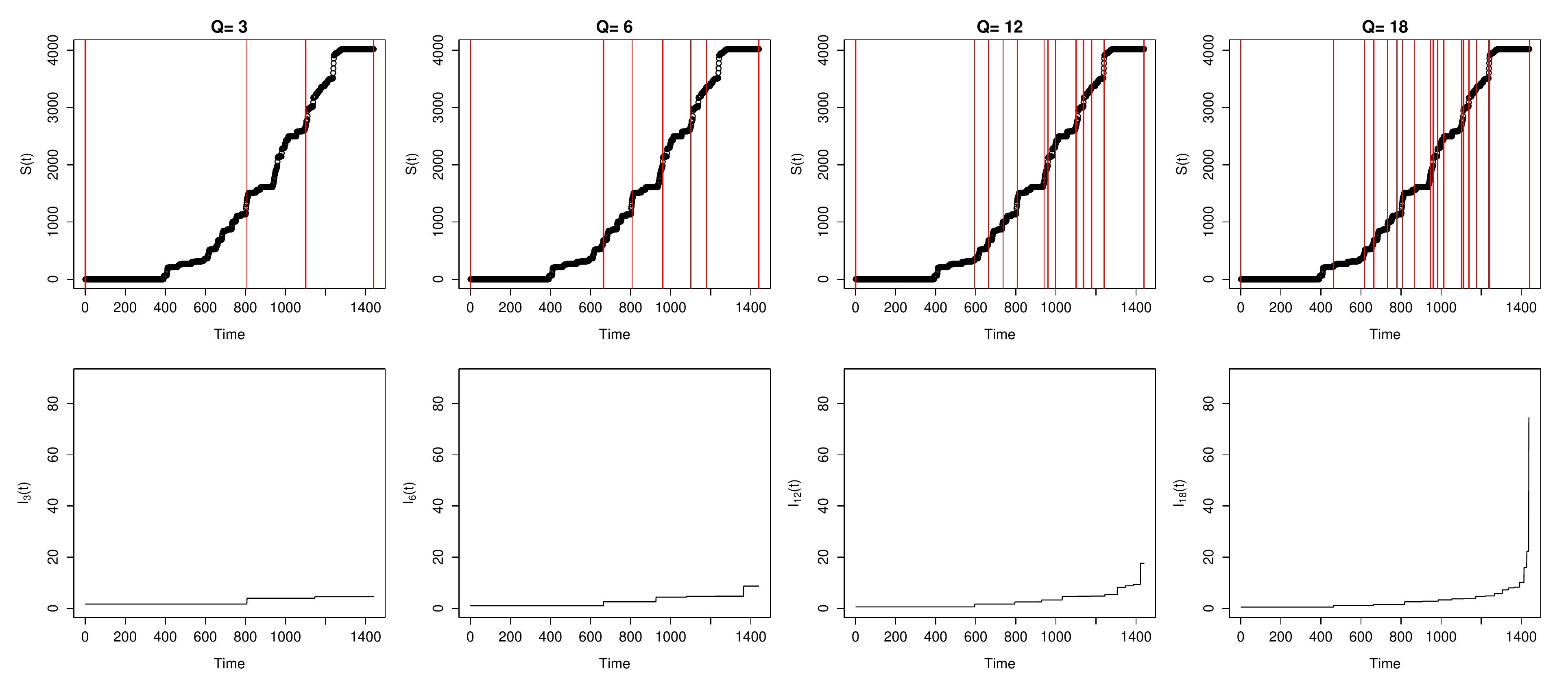}
\caption{Cumulative sum functions $S(t)$ for a specific day with $\mathcal{T} (p_0), \ldots, \mathcal{T} (p_Q)$ with $Q=3,6,12,18$, marked by red vertical lines (top row), and the corresponding quantile slope functions $I_{Q}(t)$ (bottom row).}
\label{order_K}
\end{figure}

To assess the effect of the number of quantiles $Q$, we compute $S(t)$ and $I_{Q}(t)$ on day 20684 for different values of $Q=3,6,12,18$. As shown in Figure \ref{order_K}, the $S(t)$ detects the moderate-intensity from $t=600$ to $1250$ for all $Q$'s. The magnitude of $I_{Q}(t)$ changes with an increase of $Q$. A small $Q$ makes it difficult to detect intensity information, while a large $Q$ makes the calculation time long. For the real data analysis and simulation study, we set $Q=8$, and the sensitivity test presented in Section \ref{sen_test} ensures that the clustering result is robust to the choice of $Q$ value.

\subsection{Mean Score Function}
\begin{figure}
\centering
\includegraphics[width=15cm]{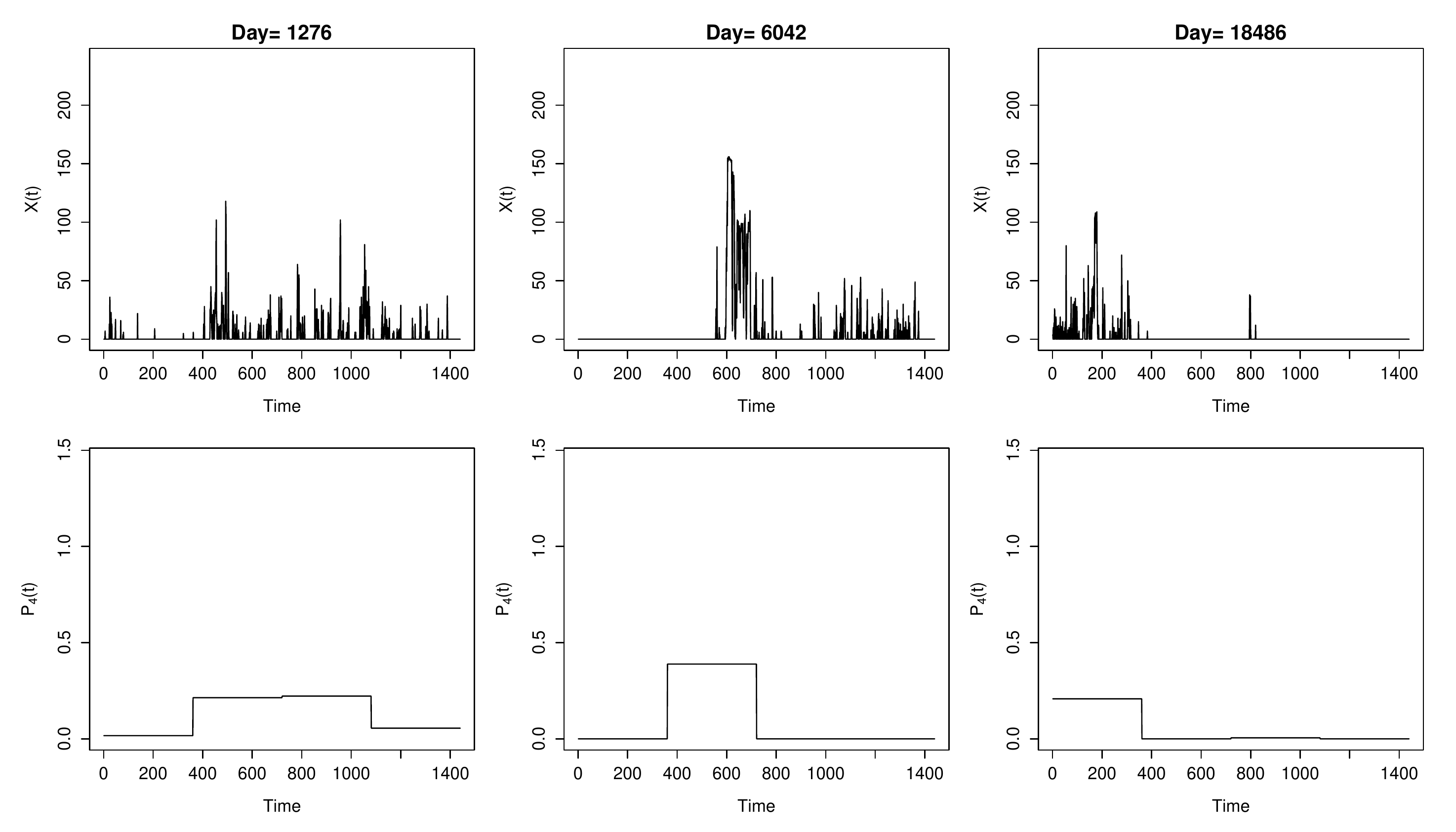}
\caption{Three step datasets $X(t)$ (top row), and the corresponding mean score functions $P_{Q} (t)$ with $Q=4$ (bottom row).}
\label{msc}
\end{figure}

We consider a new variable to reflect  the pattern of physical activity. We first define the cumulative sum of ordered steps as $S_{(t)}  := \int_0^t X_{(u)} du$ for $t \in (0, T)$, where $X_{ (t)}$ denotes the $t$th  smallest value of $\{ X{(t)}\}_{t=1}^T$. In a  similar way to the $\mathcal{T}(p)$ defined in \eqref{fp}, we define the $100p$th quantile of the ordered activity time as
$$ 
\mathcal{T}_{(p)} := \inf \{ t | S_{(t)} \geq p S_{(T)} \}, 
$$
which indicates the time when the step data reordered in ascending order achieved $100p$ of the total activity. Then, we define the score function $u(t) $ as
$$
 u (t) = q ~, ~~\text{if}~~  \mathcal{T}_{(p_q)}  < \text{rank} ( X_{(t)} ) \leq \mathcal{T}_{(p_{q+1})},
$$
where $p_q = \frac{q}{Q}$, $q=0, \ldots, Q$. Thus, the $u(t)$ represents the activity at time $t$ compared to that at other time points. For further examining the pattern of the activity, we compute the local average of $u(t)$, termed  {\em mean score function} via quantile of ordered data $X_{(t)}$ with $Q+1$ quantiles, 
 $$  
 P_{Q}(t) := \begin{cases}   \frac{1}{T/Q} \sum _{k=1}^{t_{1}} u(k) , ~~~~~~~~  \text {for} ~ 0 < t \leq t_1\\ 
	    \frac{1}{T/Q} \sum _{k= t_{1}+1}^{t_{2}} u(k)  , ~~~~ \text {for} ~ t_1 < t \leq t_2 \\ 
	   ~~~~~~~~~~~~~~~~~~~\vdots \\
 	 \frac{1}{T/Q} \sum _{k=t_{Q-1}+1}^{T} u(k) , ~~\text {for} ~ t_{Q-1} < t \leq T, \\ 
	 \end{cases}
$$ 
where $t_q= T \times p_q,$  $q=1, \ldots, Q-1$. To identify both global and local patterns of the activity, we use the local average of $u(t)$ rather than itself. Figure \ref{msc} shows the mean score function $P_{Q} (t)$ with $Q=4$ for four randomly selected days. We observe that $P_{Q} (t)$ represents the pattern of the step data, whereas the information of the amount and intensity has disappeared. For example, on day 6042, most activities occur between $t=600$ and $t=800$ and are well reflected in the corresponding mean score function. Also, on day 1276, activities are evenly distributed from $t=400$ to $t=1100$, which can be observed from the mean score function.

\section{Proposed Method for Clustering}\label{ch3}
This section proposes a clustering procedure using the variables defined in Section \ref{ch2}. For this purpose, we represent the variables defined in Section \ref{ch2} as continuous functional data on a finite-dimensional space spanned by basis functions. Let $\bX_i(t) := ( X_{1i}(t), X_{2i}(t), X_{3i}(t))^T$ for $i=1,2, \ldots , N$, $t=1, \ldots , T $, where 
\begin{itemize}
	\item $X_{1i}(t)$ -- functional data of the cumulative sum function, $S_{i}(t)$.
	\item $X_{2i}(t)$ -- functional data of the ordered quantile slope function, $I_{Q_{1},i}(t)$.
	\item $X_{3i}(t)$ -- functional data of the mean score function via quantile of ordered data, $P_{Q_{2},i}(t)$.
\end{itemize}
Here  $Q_{1}$ and  $Q_{2}$ are the predetermined numbers of quantiles used in the ordered quantile slope function and mean score function, respectively. Thus, $\bX_i(t)$ are multi-feature 
functional data that reflect the quantity, intensity, and pattern of the step counts in the $i$th day. 
 
For our analysis, we standardize the $k$th variable as 
$$
 Z_{ki}(t) := \frac{X_{ki}(t)}{\Big(\frac{1}{N}\sum _{i=1}^{N} \max\limits_{t=1,\ldots,T} X_{ki}(t)\Big)}, ~~i=1, 2,\ldots, N, ~k=1,2,3.
$$  
Then $Z_{ki}(t)$ is represented as $Z_{ki}(t) = \sum_{r=1}^{R_{k}} c_{kir}\phi_{kr}(t),~k=1,2,3,~0 \leq t \leq T,$ where $ \phi_{kr}(t)$ is the basis function for the $k$th variable, and $R_{k}$ is the number of basis functions. In this study, we use a B-spline basis with a cubic polynomial segment \citep{Boor1978}.  

We now perform a clustering procedure by applying an MFPCA to the standardized functional data $\bZ_i(t):= ( Z_{1i}(t), Z_{2i}(t), Z_{3i}(t))^T$. For a self-contained material, we briefly review the MFPCA. Suppose that we have $\bZ(t) := (Z_{1}(t), Z_{2}(t), \ldots, Z_{p}(t))^T$, $t \in \mathcal{T}$, in a Hilbert space of $p$-dimensional functions in $L_2 (\mathcal{T})$, denoted by $\mathbb{H}$. Let $\bmu (t) = (\mu_{1} (t) ,\ldots,\mu_{p} (t)) $, where $\mu_{k} (t) = E (Z_k(t) )$, $k=1, \ldots, p,$ denotes the continuous mean function, and $\bV(s,t) = \mathbb{E}[(\bZ(s)-\bmu(s))\otimes(\bZ(t)-\bmu(t))],$ $s,t \in \mathcal{T}$, denotes the  covariance matrix. The inner product is defined as 
 $$ 
 \langle \bff, \bg \rangle := \sum_{j=1}^p  \int_{\mathcal{T}} f_j(t)  g_j (t)  dt,
 $$
 where $\bff = (f_1, \ldots, f_p)^T$ and $\bg = (g_1, \ldots, g_p)^T$ in $\mathbb{H}$. The MFPCA identifies the eigenvalues and eigenfunctions that satisfy the spectral analysis of the covariance operator $C$. Specifically, we define the covariance operator $C :  \mathbb{H} \rightarrow \mathbb{H}$ on $\bff = (f_1, \ldots, f_p)^T \in \mathbb{H}$ as
$$ (Cf)^i (t) =  \sum_{j=1}^p  \int_{\mathcal{T} } V_{ij} (s,t) f_j(s)ds,$$
where $V_{ij}(s,t) $ is the $(i,j)$th element of $\bV (s,t)$. Then, by the Hilbert--Schmidt theorem \citep{Renardy2006}, there exists a complete orthogonal basis of eigenfunctions $\bpsi_{r}  = (\psi^r_1, \ldots, \psi^r_p)^T\in \mathbb{H}$ satisfying
 $$C\bpsi_{r} = \lambda_{r}\bpsi_{r}, ~~\text{for all }~{r =1, 2, \ldots }$$
and $\lambda_{r} \rightarrow 0$ as $r \rightarrow \infty$. Furthermore, the multivariate Karhunen--L\`oeve expansion of $\bZ(t)$ is
$$ \bZ(t) = \bmu(t) + \sum_{r=1}^\infty   \xi_r  \bpsi_r(t),$$
where $\xi_r:= \langle \bZ-\bmu , \bpsi_r \rangle$ is the $r$th functional principal component score. For $N$ observations, the multivariate Karhunen--L\`oeve expansion of $\bZ_i(t)$ is
$$ \bZ_i(t) = \bmu_i(t) + \sum_{r=1}^\infty   \xi_{ir}  \bpsi_r(t)~~\text{for} ~~ i=1, \ldots, N.$$
The functional principal component (FPC) scores $(\xi_{i1}, \xi_{i2}, \ldots ,\xi_{iR} )^T$ are computed as $\xi_{ir}:= \langle \bZ_i-\bmu_i , \bpsi_r \rangle$, where the number of FPCs, $R$ is determined by the proportion of the explained variance.

Finally, we apply an existing clustering method, such as $K$-means algorithm and PAM algorithm, to the FPC scores of each day.

\section{Real Data Analysis}\label{ch4}
\subsection{Data and Setup}
The step data used in this analysis are recorded for 21394 days over 79 people, with the number of days per person varying from 32 to 364. Our goal is to cluster 21394 days based on the amount, intensity, and pattern of activity. The current study focuses on clustering days, but the proposed method is readily applicable to clustering days for specific individuals with sufficient data. It can also be used to cluster individuals instead of days by connecting the daily step data from each subject as a single time series. Indeed, Section 4.4 discusses briefly clustering 79 individuals by the proposed method. 

To construct the variables, we set the number of quantiles $Q_{1}=8$ for the ordered slope function $I_{Q_1} (t)$ that is suitable to reflect intense activity, such as exercise. For the mean score function, we use $P_{Q_2}(t)$ with $Q_{2}=4$, for grouping 24-hour activity patterns: early morning (0:00-6:00), morning (6:00-12:00), afternoon (12:00-18:00), and evening (18:00-24:00). For analysis, we obtain standardized functional data, $\{ Z_{ki}(t)\}_{ i=1, \ldots, N}$, $t= 0, \ldots, T, ~k=1,2,3$, where $N$ is the number of days ($N=21394$), and $T=1440$. Note that $t=0$ and $t=1440$ correspond to 12:00 AM. 

\begin{figure}
\centering
\includegraphics[width=\linewidth]{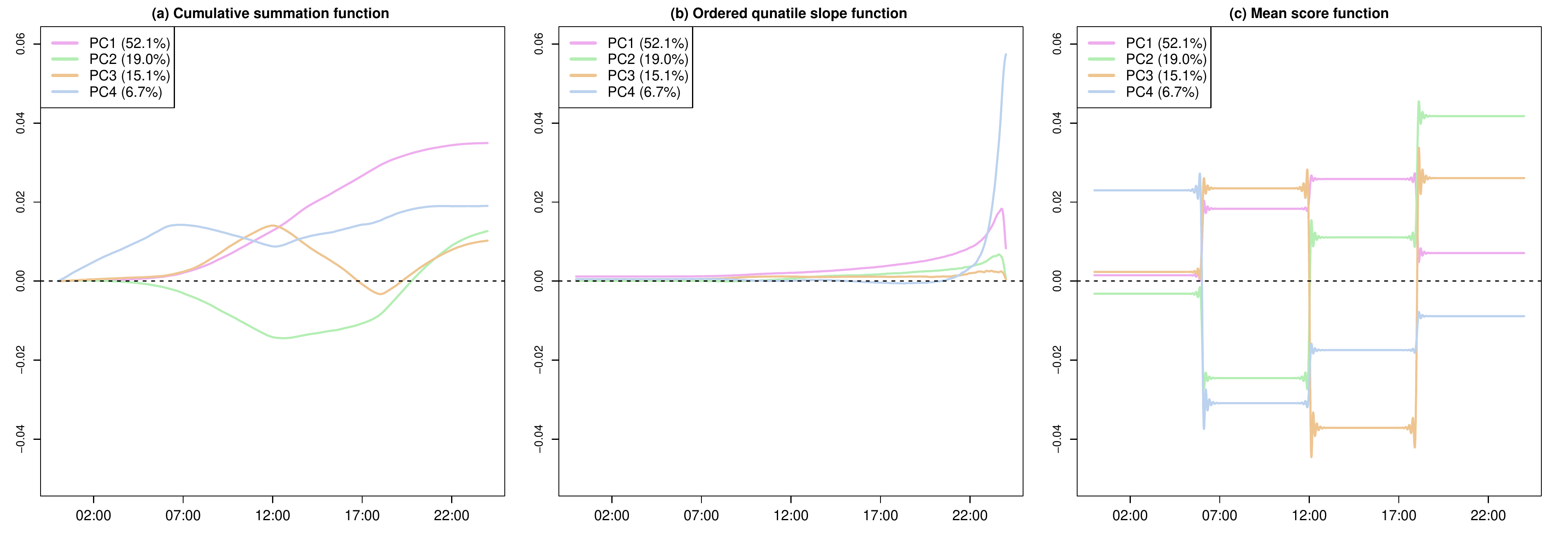}
\caption{Estimates of the four eigenfunctions for each variable; $ \psi^r_1 (t) $ (left), $\psi^r_2(t)$ (middle), and $\psi^r_3(t)$ (right), for $r=1,2,3,4$ from multivariate FPCA. The number in parenthesis indicates the percentage of the explained variance. }
\label{PCfunction}
\end{figure}

Finally, the number of FPC scores in the MFPC procedure is selected using the total explained variance. This analysis uses the four leading MFPC scores that describe 92.97\% of the total variance. Figure \ref{PCfunction} shows the four leading estimated eigenfunctions for each variable. The first two eigenfunctions of the cumulative summation function, $\psi^1_1 (t)$ and $\psi^2_1 (t)$ are contrasted with each other for 07:00-17:00, while the first three eigenfunctions of the ordered quantile slope function, $\psi^1_2 (t), \psi^2_2 (t)$, and $\psi^3_2 (t) $ look similar. Significantly different patterns in eigenfunctions of mean score function are observed.

As for the conventional clustering method applied to the MFPC scores, we use $K$-means and PAM algorithms.

\subsection{Clustering Results}

We apply the $K$-means algorithm to the MFPC scores and divide the total of 21394 days into seven subgroups. Note that for determination of an optimal number of clusters of $K$-means algorithm and PAM algorithm, we use the gap statistic \citep{Tibshirani2001}, which yields $K=7$. 

\begin{figure}
\centering
\includegraphics[width=\linewidth]{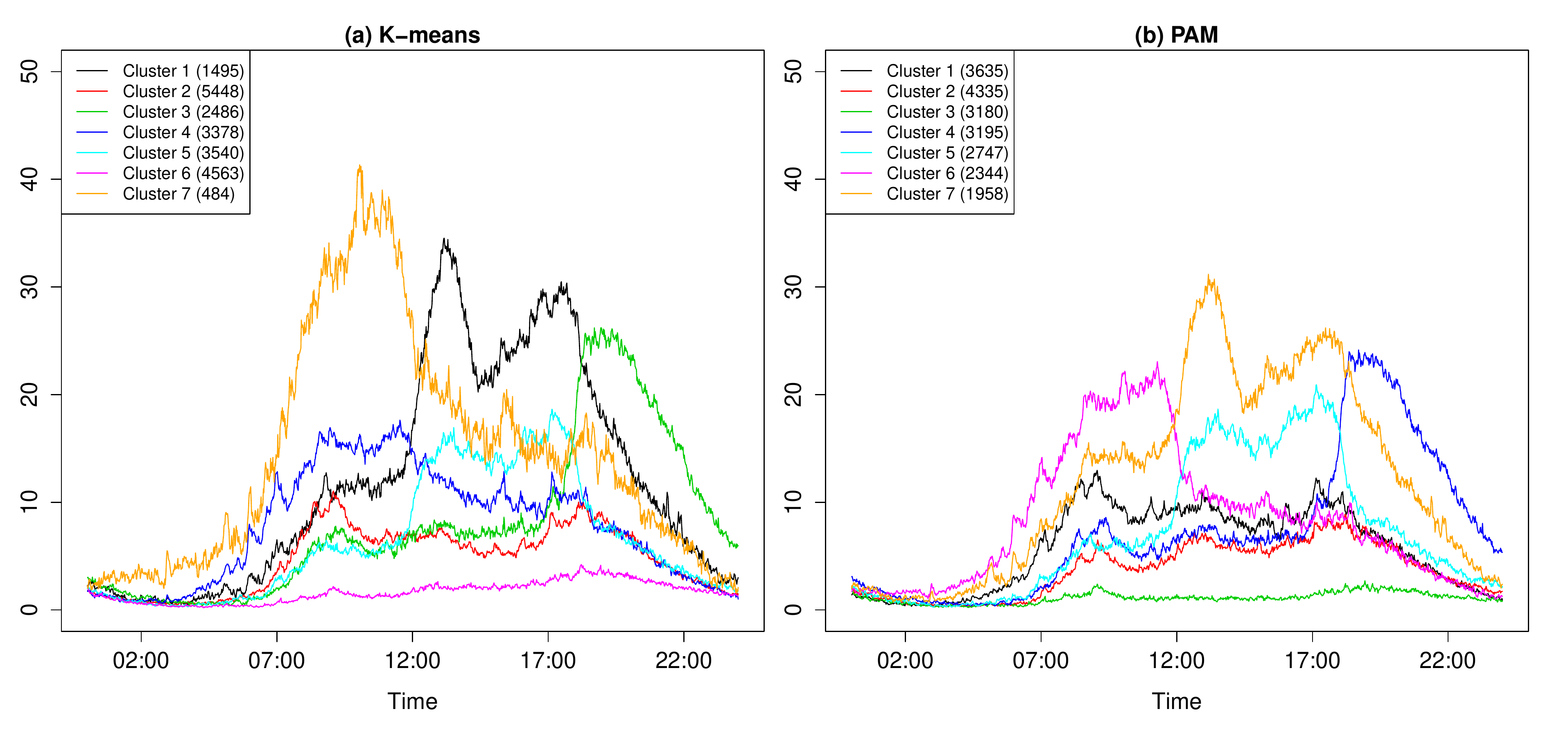}
\caption{Mean curve of step counts in each group obtained from $K$-means and PAM. The number in parenthesis indicates the number of days in each group.}
\label{kmeans_mean}
\end{figure}

Figure \ref{kmeans_mean}(a) shows the mean curve of the step data in each group. The number in parenthesis in the figure indicates the number of days belonging to each group. We make some observations: (i) Cluster 6 is identified as the lowest-quantity and lowest-intensity group. (ii) The activities in Clusters 4 and 7 appear to be concentrated in the morning, although there are some differences in the amount of activity. (iii) Clusters 1 and 5 represent activities in the afternoon, although there is more significant activity in the former than in the latter. (iv) The days belonging to Cluster 3 show active movements in the evening. (v) Finally, the days in Cluster 2 tend to be relatively constant from morning to evening, but slightly more likely during rush hour. 

\begin{figure}[!h]
\centering
\includegraphics[width=\linewidth]{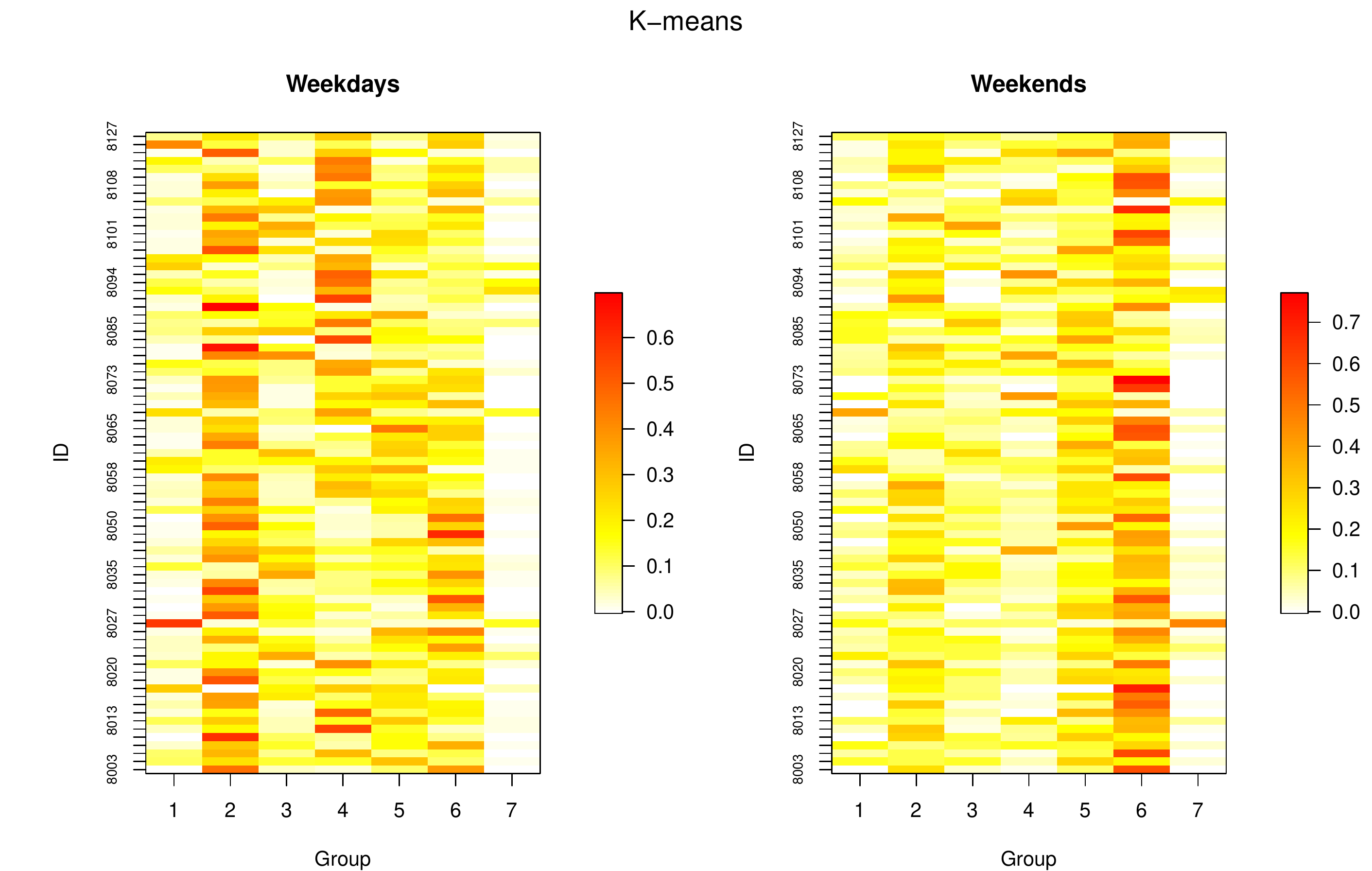}
\caption{Heatmaps of clustering results from the proposed algorithm on weekdays and weekends. The color code indicates the proportion of days that an individual belongs to each cluster.}
\label{kmeans_heatmap}
\end{figure}

The heatmaps of the clustering results are shown in Figure \ref{kmeans_heatmap} according to weekdays and weekends, indicating the proportion of days that an individual belongs to each cluster. Most individuals fall into Cluster 2 on weekdays, a group active during rush hour. Clusters 4 and 6 also include some individuals on weekdays that represent intermediate and minimum activity groups. On the other hand, most individuals belong to Cluster 6 on weekends, the minimum active group.

We now use the PAM algorithm to implement the proposed method for clustering the step data. The results are shown in Figure \ref{kmeans_mean}(b). We make some observations: (i) Clusters 3 and 7 are identified as the lowest- and the highest-quantity groups, respectively. In particular, the activities in Cluster 7 appear to be active from morning to afternoon. (ii) Cluster 4, as an intermediate-quantity group, is particularly active in the evening (18:00-24:00). (iii) The days belonging to Cluster 5 look brisk in the afternoon (12:00-18:00). (iv) The days in Cluster 6 show a lot of movement in the morning (06:00-12:00).

Compared with the results from the $K$-means algorithm, the days in each cluster are evenly distributed when using the PAM algorithm. It might be because the latter is more robust than the former. 

\begin{figure}
\centering
\includegraphics[width=\linewidth]{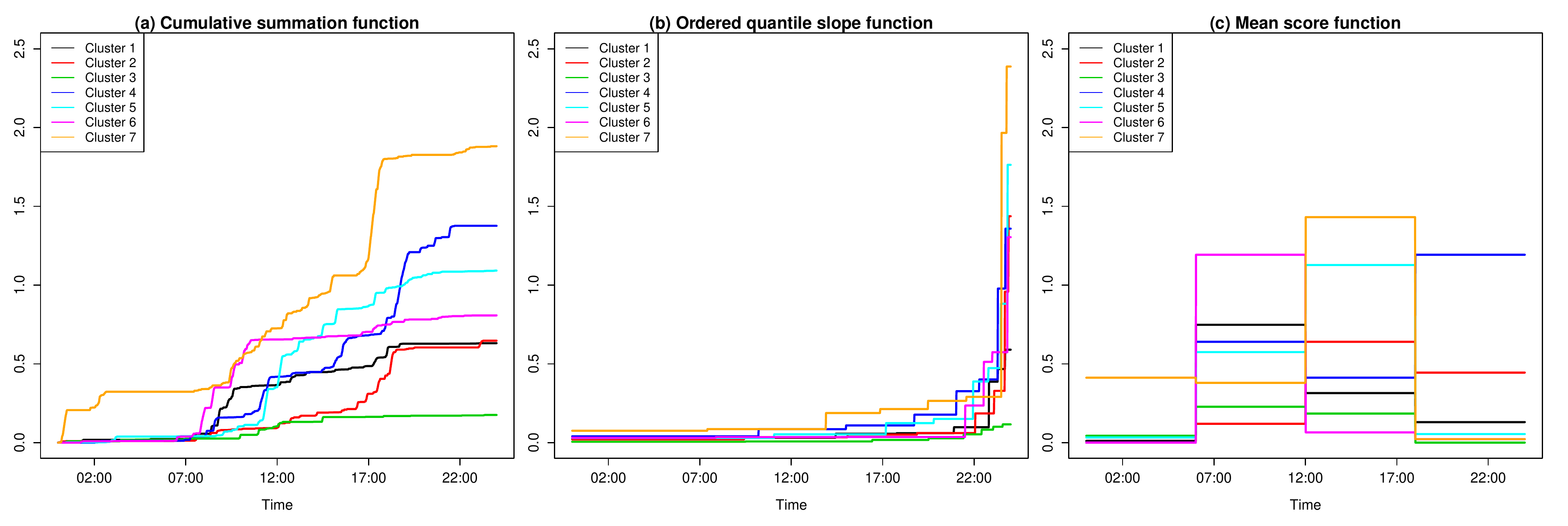}
\caption{(a) Cumulative sum functions $S(t)$, (b) ordered quantile slope functions $I_{Q_{1}}(t)$ with $Q_1 = 8$, and (c) mean score functions $P_{Q_{2}}(t)$ with $Q_2 = 4$ in each cluster obtained from PAM.}
\label{pam_amount_i_p} 
\end{figure}

For visual assessment for clustering results, we randomly select a single day from each cluster obtained by the PAM algorithm.  The corresponding cumulative sum functions, ordered quantile slope functions, and mean score functions are shown in Figure \ref{pam_amount_i_p}. From Figure \ref{pam_amount_i_p}(a), the amount of activities varies from cluster to cluster: the highest quantity group (Cluster 7), the mid-high quantity group (Clusters 4,5,6), the low-mid quantity group (Clusters 1,2), and the lowest quantity group (Cluster 3).  Figure \ref{pam_amount_i_p}(b) shows that the highest intensity is observed in Cluster 7, the middle intensity for Clusters 2,4,5,6, and the low intensity group for Clusters 1,3.  Finally, the pattern of activities is revealed from Figure \ref{pam_amount_i_p}(c) showing the average score function: the days in Clusters 1 and 6 prefer to walk in the morning (06:00-12:00), the days that belong to Clusters 2, 5, and 7 are active in the afternoon (12:00-18:00), and the movements of days in Cluster 4 are concentrated in the evening (18:00-24:00).

%

\begin{figure}
\centering
\includegraphics[width=\linewidth]{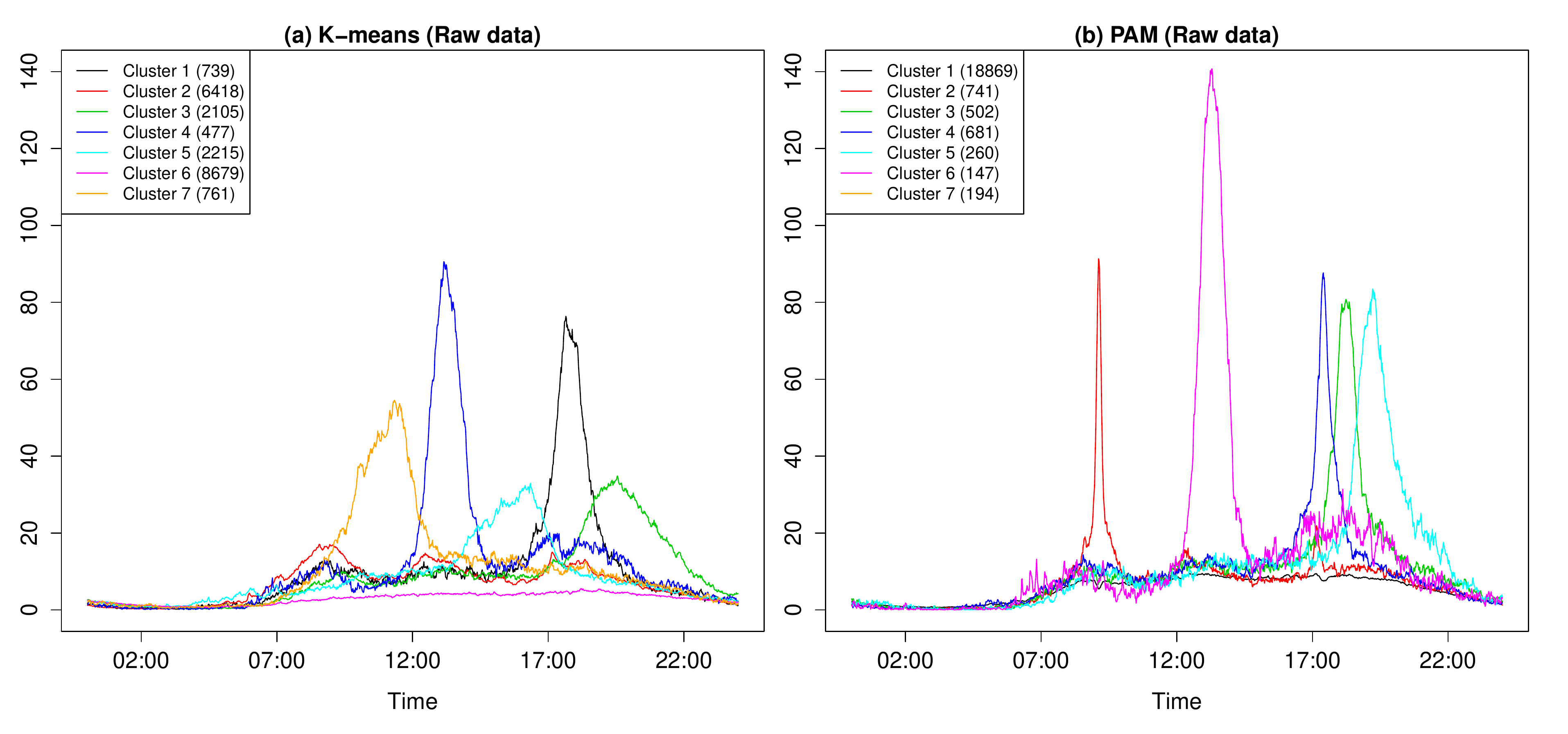}
\caption{Mean curve of step counts in each cluster obtained from (a) $K$-means and (b) PAM applied to the raw step data directly.}
\label{raw_kmeans}
\end{figure}

For further comparison, we apply the $K$-means and PAM algorithms directly to the raw step data. Figure \ref{raw_kmeans} shows the mean curve of each cluster obtained by two algorithms. It is observed that for K-mean algorithms, 70.6\% of the entire days are in Clusters 2 and 6, and for PAM algorithms, there are 88.2\% in Clusters 1. Since more than 70\% of days are clustered within a few groups, the mean curve of these groups seems flat due to the masking effect. On the other hand, from the results by the proposed method shown in Figure \ref{kmeans_mean}, we observe that days tend to be evenly distributed across clusters. We also find it difficult to observe the difference between the amount and intensity of the two clusters, although the patterns of the remaining clusters are different. 

\begin{figure}
\centering
\includegraphics[width=\linewidth]{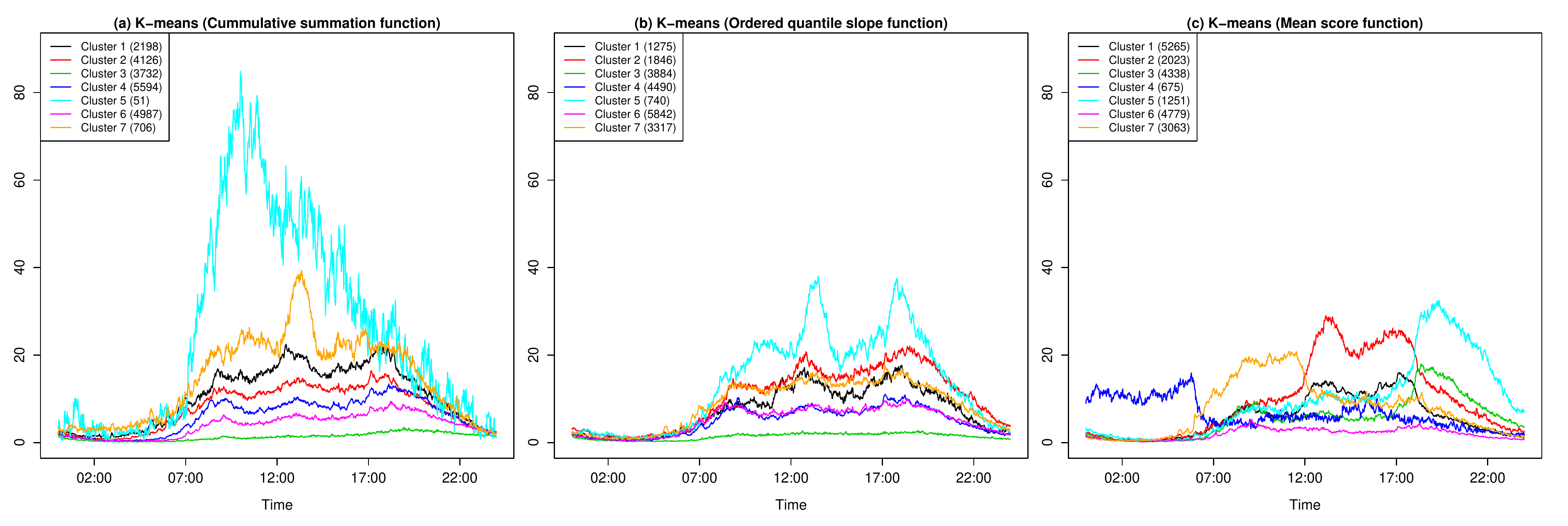}
\caption{Mean curve of step counts in each cluster obtained from $K$-means only using (a) the cumulative summation function, (b) the ordered quantile slope function and (c) the mean score function.}
\label{univariate_kmeans}
\end{figure}

The proposed three input variables, $S(t), I_{Q_1} (t), P_{Q_2} (t)$, can be used alone for $K$-means without MFPCA step, similar to \cite{Lim2019}. Figure \ref{univariate_kmeans} presents the clustering results by applying $K$-means to each input variable. As expected, the $K$-means with the cumulative summation function, $S(t)$, clusters the data according to the amount of the activity, and the result based on the mean score function, $P_{Q_2} (t)$, only reflects the pattern of the activity. We observe that the MFPCA step of the proposed method is necessary to simultaneously consider the amount, intensity, and pattern of the step data for clustering.  

%

\subsection{Sensitivity Test for $Q_{1}$ of Ordered Quantile Slope Function}  \label{sen_test}
\begin{figure}[!h]
\centering
\includegraphics[width=0.5\linewidth]{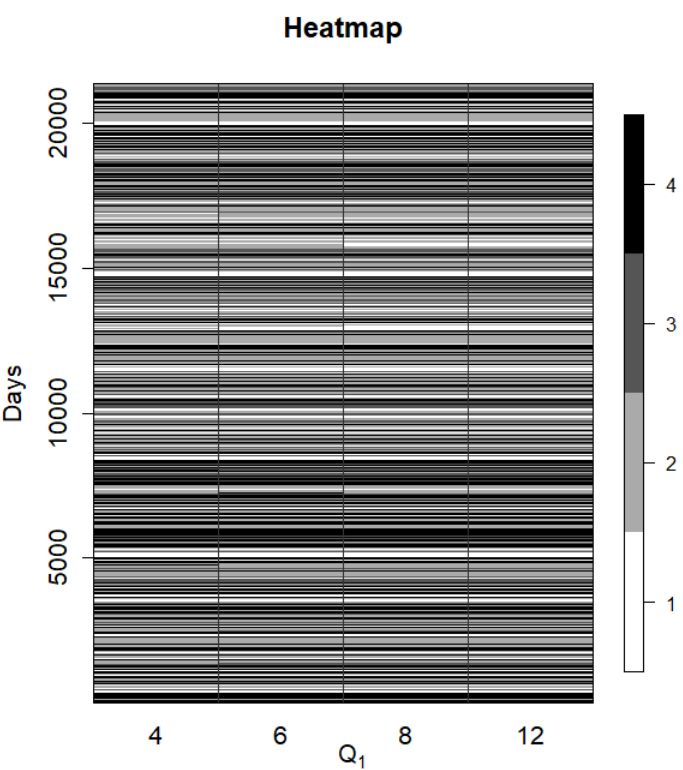}
\caption{Heatmap of clustering results obtained by the proposed method with PAM. The color code indicates the cluster groups.}
\label{heatmap_Kmeans}
\end{figure}

Unlike the $Q_{2}$ of the mean score function $P_{Q_{2}}(t)$ that can be easily set according to the time zone, it may seem quite difficult and arbitrary to select an optimal number of quantiles $Q_{1}$ of the ordered quantity function $I_{Q_{1}}(t)$. In addition, we observe from Figure \ref{order_K} that $I_{Q_{1}}(t)$ varies with the choice of $Q_1$ value. Here we perform a sensitivity test with varying values of $Q_1$. Note that the number of clusters is set to four, $K=4$. Figure \ref{heatmap_Kmeans} shows the heatmap image of clustering results obtained by PAM with $Q_{1} = 4 ,6 ,8 ,12$. It seems that the clustering result is consistent with $Q_1$ values, indicating that the proposed method is not sensitive to the number of quantiles for the ordered quantile slope function. 

\section{Simulation Study}

\subsection{Experimental Setup}

To evaluate the empirical performance of  the proposed method, we generate several simulated curves with different amounts, intensities, and patterns of activity.

\begin{figure}
\begin {center}
\includegraphics[width=145mm]{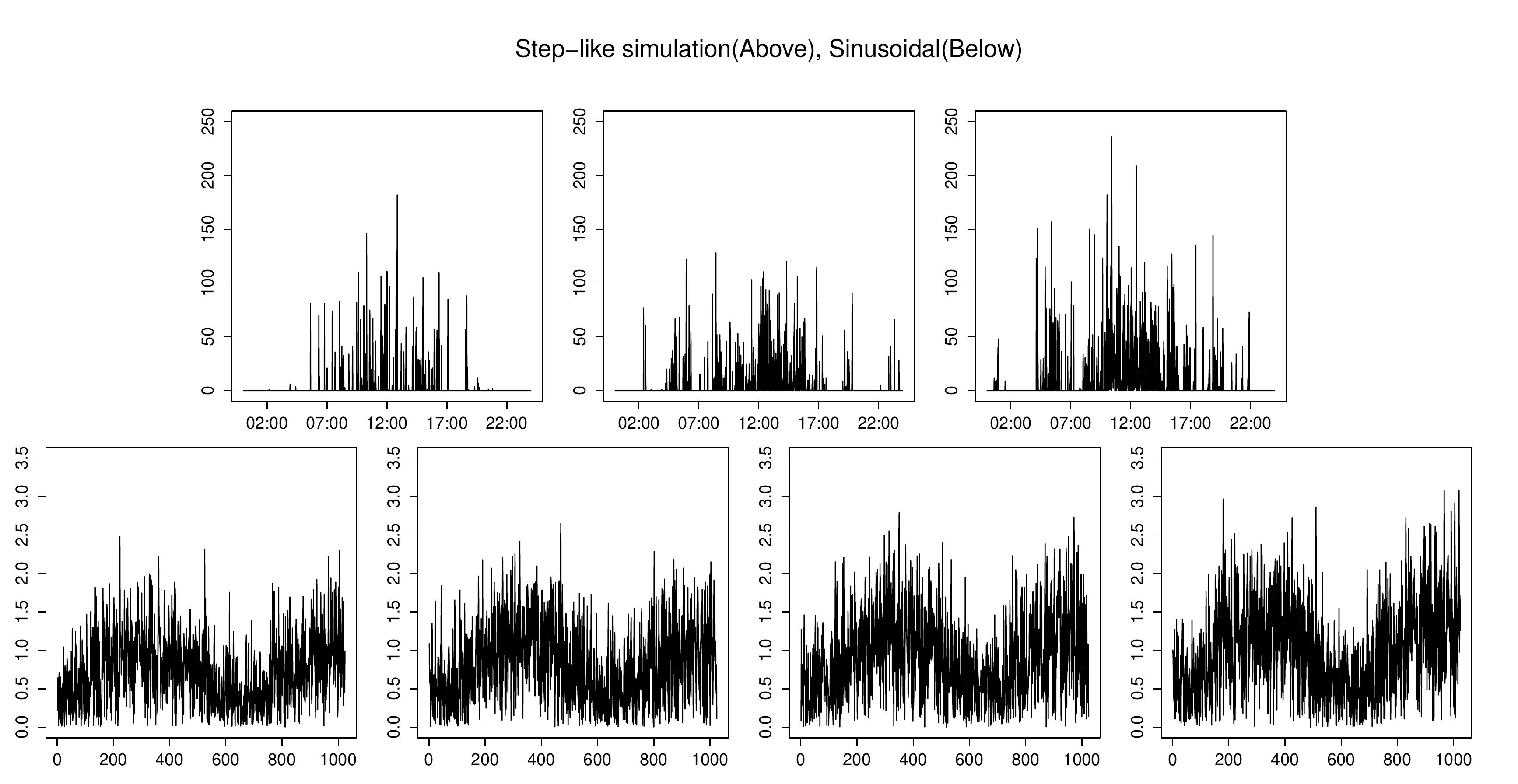} 
\caption{Simulated step-like data with different amounts (top row), and simulated sinusoidal curves with different amounts (bottom row).}
\label{sim_amount}
\end {center}
\end{figure}

\noindent 
{\underline{Curves with different amounts}}
\begin{itemize}
\item Step-like simulation data: One important feature of step data is zero-inflation. From the real step data in Section \ref{ch4}, we observe that, out of 1440 minutes, the low amount group is active for about 150 minutes, the middle amount group for about 250 minutes and the high amount group for about 350 minutes. Therefore, we generate the number of non-zero points in the $i$th curve for the $k$th group as follows:
$$
N_{i,k} = \lfloor{W}\rfloor,~~~~ W\sim {N}(\mu_k, \sigma^2),~~~i=1, \ldots, n_k,~~k=1,2,3, 
$$
where $(\mu_1, \mu_2, \mu_3)=(150,250,350),$ $\sigma^2=15$, and $N:= \sum_k n_k$. Here, $\lfloor{x}\rfloor$ denotes the largest integer less than or equal to $x$. Then, the $i$th simulated step data in group $k$, $Y_{i,k} (t)$, has a nonzero value at $t \in \mathds{T}_{i,k}$, where the number of time points in $\mathds{T}_{i,k}$ is $N_{i,k}$. To fix the intensity and pattern for all curves ($i=1, \ldots, N$), we set $\mathds{T}_{i,k}$ as follows: 75\% of $\mathds{T}_{i,k}$ are randomly located in $t=481, \ldots, 960$, and 21\% of $\mathds{T}_{i,k}$ are in $t=241, \ldots, 480, 961, \ldots, 1200$. The remaining 4\% are in $t=1, \ldots, 240$ and $t=1201, \ldots, 1440$. Now, the $i$th simulated step data in group $k$ is generated from the following exponential distribution, 
\begin{eqnarray}    
         Y_{i,k} (t) =
             \begin{cases}  \lfloor Z \rfloor, ~~Z \sim \mbox{Exp} (1/\lambda),~~~t \in \mathds{T}_{i,k},\\
       	 0, ~~~\hspace{3.2cm} t \notin \mathds{T}_{i,k},
                   \end{cases}
                   ~~ i=1, \ldots, n_k,~~k=1,2,3,
\end{eqnarray}
where $\lambda=32.5$ that denotes the estimated overall mean of the real step data. We generate $n_k=100$, $k=1,2,3$, random curves from each group. Each realization of random curves according to groups is shown in the first row of Figure \ref{sim_amount}.

\item Sinusoidal signal: We generate a random curve defined as 
\begin{eqnarray}    
              {Y_{i,k}(t_{j}) = a_k \Big|\sin(\frac{5t_{j}}{T})+\varepsilon_{ijk} \Big|,~~~j=1, \ldots, T, ~i=1, \ldots, n_k, ~k=1,\ldots, 4},
                \label{sin}
\end{eqnarray}  
where $t_{j}=\frac{j-1}{T}$ with $T=1024$, and $\varepsilon_{ijk} \sim N(0,\sigma^{2})$ with $\sigma^{2}=0.5$. Then, we set $\ba=(a_1, a_2, a_3, a_4)=( 1,1.1,1.2,1.3) $ to reflect the difference in the amounts. Here we generate $n_k=50$ random curves in each group. Sample curves from each group are shown in the second row of Figure \ref{sim_amount}.
\end{itemize}
    
\begin{figure}
\begin {center}
\includegraphics[width=135mm]{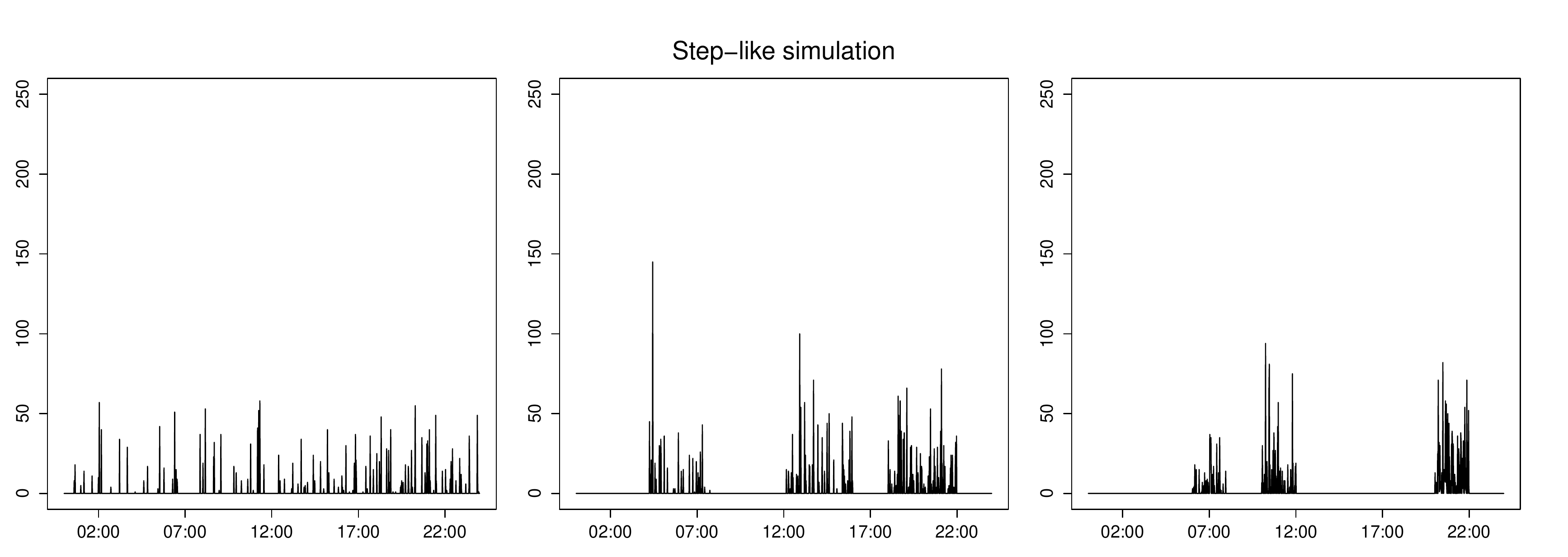}
\caption{Sample step-like data with different intensity.}
\label{sim_intensity}
\end {center}
\end{figure}

\noindent 
{\underline{Curves with different intensity}}
\begin{itemize}
\item Step-like simulation data: We generate $n_k=100$ curves with different intensities according to three groups ($k=1,2,3$). Similarly, the number of nonzero points in the $i$th curve for the $k$th group is generated as
                      \begin{eqnarray}    
                      N_{i,k} = \lfloor{W}\rfloor,~~~~ W\sim {N}(\mu, \sigma^2),~~~i=1, \ldots, n_k,~~k=1,2,3, 
                      \label{step1}
                          \end{eqnarray}
where $\mu = 150$ and $\sigma^2 = 10$. Then, the $i$th simulated step data in group $k$ have nonzero values at $t \in \mathds{T}_{i,k}$, where the number of time points in $\mathds{T}_{i,k}$ is $N_{i,k}$.
        
To further vary the intensities in the data, we define $\mathds{T}_{i,k}$ differently for each group $k$. For the first group, we generate curves with low intensity as follows: 20\% of $\mathds{T}_{i,1}$ are randomly located in $t=1, \ldots, 480$, and 30\% and 50\% of $\mathds{T}_{i,1}$ are in $t=481, \ldots,  960$ and $t=961, \ldots, 1440$, respectively. For the second group, we define $\mathds{T}_{i,2}$ with a narrower interval than that of $\mathds{T}_{i,1}$: 20\% of $\mathds{T}_{i,2}$ are randomly located in one of two intervals, $t=1, \ldots, 240$ or $t=241, \ldots, 480$; 30\% of $\mathds{T}_{i,2}$ are randomly located in one of $t=481, \ldots, 720$ and $t=721, \ldots, 960$; and 50\% of $\mathds{T}_{i,2}$ are randomly located in small intervals in $t=961, \ldots, 1440$. For the last group, we generate high-intensity curves: 20\% of $\mathds{T}_{i,3}$ are randomly located in one of the four intervals, $t=1, \ldots, 120$, $t=121, \ldots, 240$, $t=241, \ldots, 360$, and $t=361, \ldots, 480$;  30\% of $\mathds{T}_{i,3}$ are randomly located in one of the four intervals, $t=481, \ldots, 600$, $t=601, \ldots, 720$, $t=721, \ldots, 840$, and $t=841, \ldots, 960$, and 50\% of $\mathds{T}_{i,3}$ are densely located in small intervals in $t=961, \ldots, 1440$. Now, the $i$th simulated step data in group $k$ are defined as
                  \begin{eqnarray}    
         Y_{i,k} (t) =
             \begin{cases} \lfloor Z \rfloor, ~~Z \sim \mbox{Exp} (1/\lambda),~~~t \in \mathds{T}_{i,k},\\
       	 0, ~~~\hspace{3.2cm} t \notin \mathds{T}_{i,k},
                   \end{cases}
                   ~~ i=1, \ldots, n_k,~~k=1,2,3,
                   \label{step2}
                \end{eqnarray}
                where $\lambda=20$.                
Three sample curves are shown in Figure \ref{sim_intensity}.
       
    \end{itemize}
    
\begin{figure}
\begin {center}
\includegraphics[width=135mm]{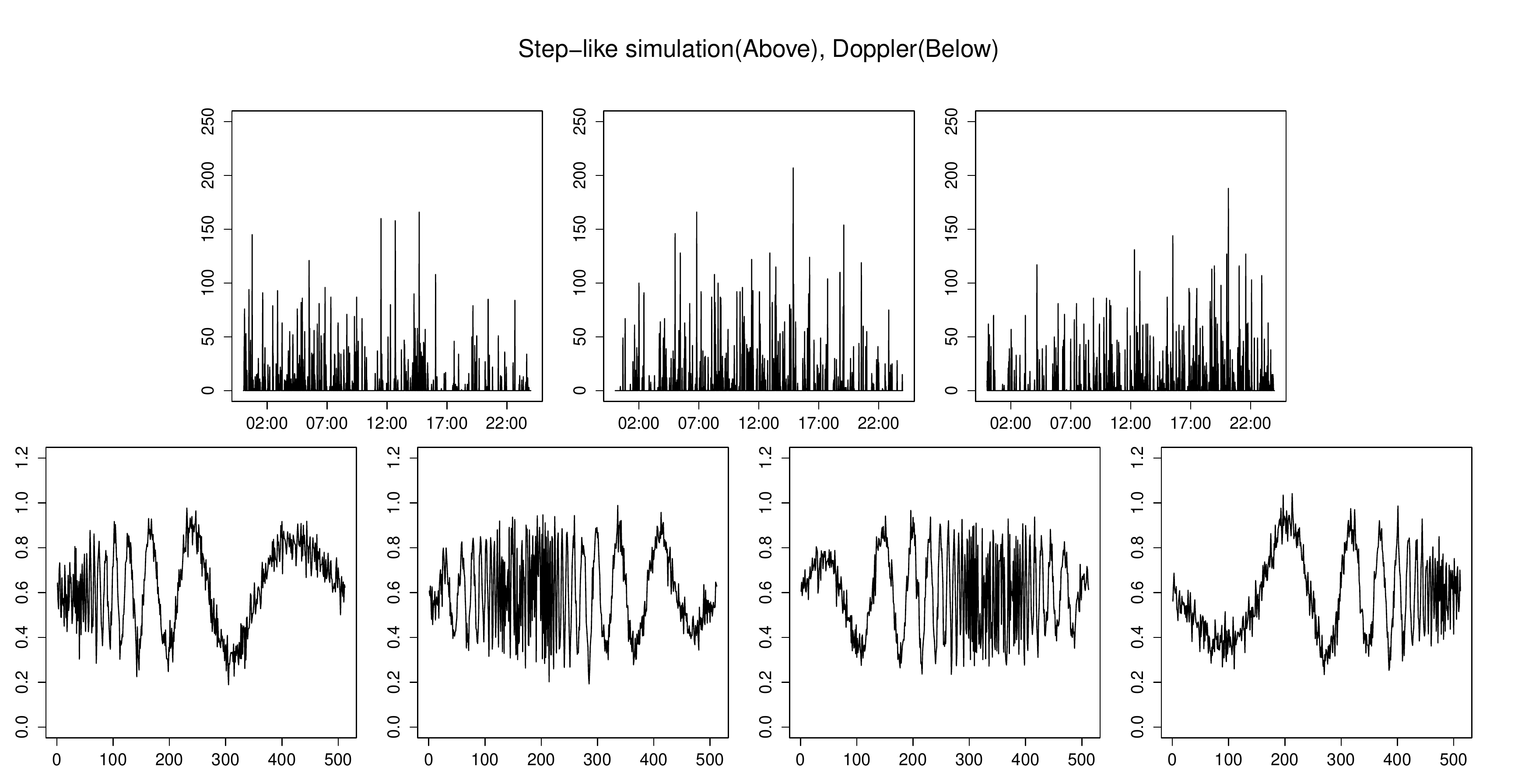}
\caption{Sample step-like data with different patterns (top), and sample doppler curves with different patterns (bottom).}
\label{sim_pattern}
\end {center}
\end{figure}

\noindent 
{\underline{Curves with different patterns}} 
\begin{itemize}
\item Step-like simulation data: We generate $n_k=100$ random curves with different patterns from three groups $(k=1,2,3)$.   We generate $N_{i,k}$ as \eqref{step1} with $\mu = 250$ and $\sigma^2= 15$, and $ Y_{i,k} (t)$ is generated as \eqref{step2} with $\lambda=32.5$. To have a different pattern for each group, we generate $\mathds{T}_{i,k}$ differently for $k=1,2,3$. For the first group ($k=1$), 45\% of $\mathds{T}_{i,k}$ are randomly located in $t=1, \ldots, 480$, and 35\% and 20\% of $\mathds{T}_{i,k}$ are in $t=481, \ldots,  960$ and $t=961, \ldots, 1440$, respectively. For the second group ($k=2$), 35\% of $\mathds{T}_{i,k}$ are randomly located in $t=1, \ldots, 480$, and 45\% and 20\% of $\mathds{T}_{i,k}$ are in $t=481, \ldots,  960$ and $t=961, \ldots, 1440$, respectively. For the last group ($k=3$), the proportions are 20\%, 35\%, and 45\%, respectively. Sample curves from each group are plotted in the first row of Figure \ref{sim_pattern}.            
           
\item Shifted Doppler signal: We generate 50 random curves in four groups that have different patterns:
$$
Y_{i,k}(t_{j})=0.6+0.6\sqrt{t_{j}(1-t_{j})}~\mathrm{sin}\Big(\frac{2.1\pi}{t_{j}-t_{0,k}}\Big) + \epsilon_{ijk},~~j=1, \ldots, T, ~i=1, \ldots, n_k, ~k=1,\ldots, 4,
$$
where $t_{j} = \frac{j-1}{T}$, $T=512$ and $\epsilon_{ijk} \sim N(0, 0.05^2)$. To have a different pattern for each group, we set the shift parameter $t_{0,k}=0,1/3,2/3,1$ for each $k$. Sample curves from each group are plotted in the last row in Figure \ref{sim_pattern}.      
\end{itemize}

We compare the proposed methods with two existing methods used for clustering multivariate functional data, FunFEM and FunHDDC. 
\begin{itemize}
    \item MFPCA-Kmeans: Proposed method with the K-means algorithm.
    \item MFPCA-PAM: Proposed method with the PAM algorithm.
    \item FunFEM: Functional clustering based on discriminative functional mixture modeling of \cite{Bouveyron2016}.
    \item FunHDDC: Functional clustering based on functional latent mixture modeling of \cite{Schmutz2020}.
\end{itemize}
For the proposed methods, we set the number of quantiles for the ordered quantile slope function $I_{Q_1}(t)$ as $Q_1=8$ and the number of quantiles for the mean score function $P_{Q_2}(t)$ as $Q_2=4$. 

\subsection{Results}

As for evaluation measure, we use the correct classification rate ({\tt CCR}) ($\%$) and the adjusted Rand index ({\tt aRand}) of \cite{Hubert1985}. Note that {\tt aRand} is a corrected version of the Rand index \citep{rand1971objective} that measures the correspondence between two partitions classifying the object pairs in a contingency table. It further adjusts the Rand index to have an expected value of zero with an upper bound. Thus, a larger aRand value indicates a higher similarity between the two partitions.  

The CCR (\%) and aRand values over 100 simulations are listed in Table \ref{t1}. We have some observations: (i) The proposed methods work well in almost every case. (ii) For the curves with different amounts, the proposed methods outperform two existing methods. (iii) The proposed MFPCA-PAM outperforms for clustering curves with different patterns. (iv) For the intensity cases, the MFPCA-PAM provides the best results.

\begin{table}
\footnotesize
\caption{Means and standard deviations (in parentheses) of the correct classification rate (CCR) and adjusted rand index (aRand) values.}
\begin{center}
\begin{tabular}{c c  c  c  c c  }
    \hline
\multirow{2}{*}{test function}  & \multirow{2}{*}{$k$ } &   \multicolumn{4}{c}{CCR} \\ \cline{3-6}
    &  &   MFPCA-Kmeans &  MFPCA-PAM & FunFEM & FunHDDC   \\ \hline
{\bf Amount } &     & &&&  \\ 
 Step simulation & 3  & 0.9911 (0.005)& {\bf 0.9914 (0.006)} & 0.6440 (0.026) &0.6210 (0.083)  \\
 Sinusoidal & 4 & 0.9225 (0.134) & {\bf 0.9613 (0.016)} & 0.6312 (0.073)& 0.9347 (0.117)  \\   \hline
 {\bf Pattern }&    & &&& \\ 
Step simulation & 3  & 0.8636 (0.215) & {\bf 0.9944 (0.003)} & 0.7279 (0.046) & 0.7158 (0.101) \\
 Doppler & 4 & 0.8244 (0.184) & 0.9767 (0.010) & {\bf 1 (0)} & 0.8269 (0.171)\\   \hline
{\bf  Intensity }&     & &&& \\ 
 Step simulation & 3  & {\bf 0.6298 (0.037)} & 0.6021 (0.052) & 0.5972 (0.070) & 0.5993 (0.075)  \\  \hline
&&& \\
\hline
\multirow{2}{*}{test function}  & \multirow{2}{*}{$k$ } &   \multicolumn{4}{c}{aRand} \\ \cline{3-6}
    &  &   MFPCA-Kmeans&  MFPCA-PAM & FunFEM & FunHDDC   \\ \hline
{\bf Amount } &    & &&& \\ 
 Step simulation & 3  &   0.9735 (0.016) & {\bf 0.9743 (0.016)} & 0.3826 (0.041) &0.3428 (0.1277) \\
 Sinusoidal & 4  & 0.8825 (0.161) & {\bf 0.9008 (0.040)} & 0.4415 (0.063) &  0.8985 (0.129) \\   \hline
 {\bf Pattern }&   & &&& \\ 
Step simulation & 3   & 0.8397 (0.252) &{\bf 0.9832 (0.010)}& 0.4881 (0.077) & 0.5472 (0.086) \\
 Doppler & 4 & 0.7956 (0.187) & 0.9405 (0.025) & {\bf 1 (0)} & 0.8105 (0.188) \\   \hline
{\bf  Intensity }&  & &&& \\ 
 Step simulation & 3 & {\bf 0.3440 (0.054)} & 0.2548 (0.079) & 0.2796 (0.1225) & 0.2886 (0.1276) \\  \hline

\end{tabular}
\end{center}
\label{t1}
\end{table}


\afterpage{\clearpage}
\section{Conclusion}
In this paper, a new clustering method is proposed for discrete, high-dimensional, and zero-inflated step data. We introduce new variables that reflect the unique characteristics of the data while maintaining important information from the original data. By applying the MFPCA-based method to the new variables, we can simultaneously account for the multiple features--amount, intensity, and pattern---of step data in the clustering algorithm. Through numerical experiments involving a simulation study and real data analysis, the proposed method shows efficient clustering performance of various functional data, including step count data. We believe that our study contributes to the literature by greatly expanding the range of multivariate function data clustering. Finally, it is necessary to determine some parameters, such as the optimal number of quantiles $Q$, to implement the proposed method. It is left for future work.

\afterpage{\clearpage}


\begin{thebibliography} {}

\bibitem[{Balli et~al.(2005)}]{Balli2017}
    Balli, S., Sa\u{g}bas, E. A. and Hokimoto, T. (2017). The usage of statistical learning methods on wearable devices and a case study: activity recognition on smartwatches. {\it Advances in Statistical Methodologies and Their Application to Real Problems}, InTech Press, Rijeka, 259--277. 

\bibitem[{Bassett et~al.(2010)}]{Bassett2010}
    Bassett Jr, D. R., Wyatt, H. R., Thompson, H., Peters, J. C. and Hill, J. O. (2010). Pedometer-measured physical activity and health behaviors in United States adults. {\it Medicine \& Science in Sports \& Exercise}, {\bf 42}, 1819.

\bibitem[{Bouveyron et~al.(2016)}]{Bouveyron2016}
    Bouveyron, C., Come, E. and Jacques, J. (2016). The discriminative functional mixture model for the analysis of bike sharing systems. {\it Annals of Applied Statistics}, {\bf 9}, 1726--1760. 

\bibitem[{Cheung et~al.(2018)}]{Cheung2018}
    Cheung, Y. K., Hsueh, P. Y. S., Ensari, I., Willey, J. Z., and Diaz, K. M. (2018).
    Quantile coarsening analysis of high-volume wearable activity data in a longitudinal observational study.  {\it Sensors}, {\bf 18,} 3056. 
    

\bibitem[{Chiou and Li(2007)}]{Chiou2007}
    Chiou, J. M. and Li, P. L. (2007). Functional clustering and identifying substructures of longitudinal data. {\it Journal of the Royal Statistical Society Series B}, {\bf 69}, 679--699.

\bibitem[{Chiou et~al.(2014)}]{Chiou2014}
    Chiou, J. M., Chen, Y. T. and Yang, Y. F. (2014). Multivariate functional principal component analysis: A normalization approach. {\it Statistica Sinica}, {\bf 24}, 1571--1596.
    
\bibitem[{de Boor(1978)}]{Boor1978}
    de Boor, C. (1978). {\it A Practical Guide to Splines}. Springer-Verlag, New York. 

\bibitem[{Hubert and Arabie(1985)}]{Hubert1985}
    Hubert, L. and Arabie, P. Comparing partitions. (1985). Comparing partitions. {\it Journal of Classification}, {\bf 2}, 193--218. 

\bibitem[{Jacques and Preda(2014)}]{Jacques2014}
    Jacques, J. and Preda, C. (2014). Model-based clustering for multivariate functional data. {\it Computational Statistics and Data Analysis}, {\bf 71}, 92--106.

\bibitem[{Kaufman and Rousseeuw(1987)}]{Kaufman1987}
    Kaufman, L. and Rousseeuw, P.J. (1987). Clustering by means of medoids. {\it Statistical Data Analysis Based on the $L_1$-Norm and Related Methods}, North-Holland, 405--416.
    
\bibitem[{Kaufman and Rousseeuw(1990)}]{Kaufman1990}
Kaufman, L. and Rousseeuw, P. (1990). {\it Finding Groups in Data: An Introduction to Cluster Analysis}. Wiley, New York. 

\bibitem[{Le Masurier et~al.(2005)}]{Masurier2005}
    Le Masurier, G.C., Beighle, A., Corbin, C.B., Darst, P.W., Morgan, C., Pangrazi, R.P., Wilde, B. and Vincent, S.D. (2005). Pedometer-determined physical activity levels of youth. {\it Journal of Physical Activity and Health}, {\bf 2}, 159--168.

\bibitem[{Lim et~al.(2019)}]{Lim2019}
    Lim, Y., Oh, H.-S.  and Cheung, K. (2019). Functional clustering of accelerometer data via transformed input variables. {\it Journal of the Royal Statistical Society Series C}, {\bf 68}, 495-520.

\bibitem[{Ramsay and Silverman(2005)}]{Ramsay2005}
    Ramsay, J.O. and Silverman, B.W. (2005). {\it Functional Data Analysis}, Second edition. Springer, New York.

\bibitem[{Rand(1971)}]{rand1971objective} 
    Rand, W. M. (1971). Objective criteria for the evaluation of clustering methods. {\it Journal of the American Statistical Association}, {\bf 66}, 846--850.  

\bibitem[{Renardy(2006)}]{Renardy2006} 
    Renardy, M. and Rogers, R. C. (2006). {\it An Introduction to Partial Differential Equations},  Springer, New York.

\bibitem[{Schmutz et~al.(2020)}]{Schmutz2020}
    Schmutz, A., Jacques, J., Bouveyron, C., Cheze, L. and Martin, P. (2020). Clustering multivariate functional data in group-specific functional subspaces. {\it Advances in Data Analysis and Classification}, In Press.

\bibitem[{Shoaib et~al.(2015)}]{Shoaib2015}
    Shoaib, M., Bosch, S., Incel, O., Scholten, H. and Havinga, P. (2015). A survey of online activity recognition using mobile phones. {\it Sensors}, {\bf 15}, 2059--2085.

\bibitem[{Tibshirani et~al.(2001)}]{Tibshirani2001}
    Tibshirani, R., Walther, G. and Hastie, T. (2001). Estimating the number of clusters in a data set via the gap statistic. {\it Journal of the Royal Statistical Society Series B}, {\bf 63}, 411--423.




\end{thebibliography}
\end{document}